\documentclass[prb,two column,floats,showpacs,amsmath,amssymb]{revtex4}

\usepackage{graphicx}
\usepackage{array}
\usepackage{hyperref}

\usepackage{color}

\usepackage{dsfont}

\usepackage{amsmath}
\usepackage{amssymb}

\newcommand{\I}{\mathrm{i}}

\DeclareMathOperator{\Tr}{Tr}
\newcommand{\tr}[1]{\Tr[ #1 ]}

\newcommand{\av}[1]{\langle #1 \rangle}
\newcommand{\bra}[1]{\langle #1 |}
\newcommand{\ket}[1]{| #1 \rangle}

\newcommand{\mymod}[2]{#1\bmod#2}

\newcommand{\oH}{\hat{H}}
\newcommand{\oS}{\hat{S}}
\newcommand{\oI}{\hat{I}}
\newcommand{\oJ}{\hat{J}}

\newcommand{\oF}{\hat{F}}

\newcommand{\tauzero}{\tau_{0}}
\newcommand{\Tzero}{T_{0}}
\newcommand{\tauone}{\tau_{1}}
\newcommand{\Tone}{T_{1}}

\newcommand{\ZE}{\mathrm{ZE}}
\newcommand{\HI}{\mathrm{HI}}
\newcommand{\mymax}{\mathrm{max}}
\newcommand{\mymin}{\mathrm{min}}

\begin{document}

% -------------------------------------------------------------------------------------------------------------------------------------------
% document settings
% -------------------------------------------------------------------------------------------------------------------------------------------

% -------------------------------------------------------------------------------------------------------------------------------------------
% beginning of text-body
% -------------------------------------------------------------------------------------------------------------------------------------------

\title{Thermal electron spin flip in quantum dots}
\author{Moritz Fuchs}
\affiliation{Institut f\"{u}r Theoretische Physik und Astrophysik,
Universit\"{a}t W\"urzburg, D-97074 W\"urzburg, Germany}

\author{Felix Krau\ss}
\affiliation{Institut f\"{u}r Theoretische Physik und Astrophysik,
Universit\"{a}t W\"urzburg, D-97074 W\"urzburg, Germany}

\author{Daniel Hetterich}
\affiliation{Institut f\"{u}r Theoretische Physik und Astrophysik,
Universit\"{a}t W\"urzburg, D-97074 W\"urzburg, Germany}

\author{Bj\"{o}rn Trauzettel}
\affiliation{Institut f\"{u}r Theoretische Physik und Astrophysik,
Universit\"{a}t W\"urzburg, D-97074 W\"urzburg, Germany}

\date{\today}

\begin{abstract}
We study a thermally induced spin flip of an electron spin located in a semiconductor quantum dot. This interesting effect arises from an intriguing interplay between the Zeeman coupling to an external magnetic field and the hyperfine interaction with the surrounding nuclear spins. By considering a minimal model, we explain the main mechanism driving this spin flip and analyze its dependence on the strength of the external magnetic field, the number of nuclear spins and the ratio of the electron and nuclear Zeeman energies, respectively. Finally we show, that this minimal model can be applied to experimentally relevant QDs in III-V heterostructures, where we explicitly predict the temperature at which the spin flip occurs.
\end{abstract}


\maketitle

\section{Introduction}
\label{sec:I}

The physics of an electron spin confined to a quantum dot (QD) has been in the focus of condensed matter research for many years due to its possible applications in quantum information processing. Following the seminal work\cite{Loss1998} of Loss and \mbox{DiVincenzo}, the two eigenstates of the electron spin can be used to define a quantum bit (qubit). However, in the host nanostructures, the electron spin is subject to a variety of different interactions\cite{Schliemann2003,Coish2009} leading to a loss of the stored information and decoherence. Important examples for these interactions are the hyperfine interaction (HI) with surrounding nuclear spins or the coupling of the electron spin to electrical fields via spin orbit interaction. The dynamics of the electron spin influenced by these effects - alone or combined - have been studied in great detail\cite{Khaetskii2002,Coish2004,Petta2005,Witzel2005,Klauser2006,Koppens2008,Fischer2008,Greilich2009,Eble2009,Cywinski2009a,Bluhm2010} in the last years enhancing our knowledge about spin physics in QDs\cite{Hanson2007,Urbaszek2013,Kloeffel2013}. Along with these insights also strategies were developed to reduce the loss of information\cite{Chekhovich2013}. However, while electrical noise seems fairly controllable, the HI with the nuclear spins remains nonetheless as a major source of decoherence, since nuclear spins are intrinsic to most QD host materials such as the widely used III-V heterostructures. Yet recently, also a change of paradigm can be observed, where the unavoidable interaction with the nuclear spins is considered rather as a resource for interesting physics than an obstacle.\cite{Ribeiro2010} 

In this notion, we want to present our findings on a thermally induced flip of the electron spin. These results were obtained by applying standard statistical physics to a minimal model for spin dynamics in a QD consisting of an external magnetic field to which the electron and the nuclear spins couple and the HI which links the electron spin to all nuclear spins. The effect of an external magnetic field on the electron is described by the Zeeman Hamiltonian
\begin{equation}
\oH^{S}_{\ZE}
=
g^{*} \mu_{B} B_{z} \oS_{z}
\,,
\label{eq:HZES}
\end{equation}
where $g^{*}$ is the effective g-factor\cite{Winkler2003} of the electron, $\mu_{B}$ is the Bohr magneton, and $\oS_{z}$ is the electron spin component parallel to the external magnetic field $B_{z}$. Likewise, nuclear spins also align with respect to the external magnetic field by means of another Zeeman term
\begin{equation}
\oH^{I}_{\ZE}
=
\sum_{k=1}^{K} \hat{h}_{\ZE}^{k}
=
-\sum_{k=1}^{K}\, g_{N} \mu_{N} B_{z} \oI_{k,z}
\,,
\label{eq:HZEI}
\end{equation}
where $g_{N}$ is the nuclear g-factor, $\mu_{N}$ is the nuclear magneton and the sum is running over all nuclear spin $z$ components $\oI_{k,z}$. Note the relative sign difference between Eq.~(\ref{eq:HZES}) and Eq.~(\ref{eq:HZEI}), which resembles the negative sign of the electron charge.

By writing the nuclear Zeeman Hamiltonian in the form of Eq.~(\ref{eq:HZEI}), we already assume that there is only one spin species present in the dot. This assumption will simplify our reasoning and, thus, allows us to better identify the main physics being relevant for our results. Distinguishing different spin species would not change our findings significantly as we discuss in the end of this article. This specific choice also simplifies the HI Hamiltonian
\begin{equation}
\oH_{\HI}
=
A_{\HI}  \sum_{k=1}^{K}\, |\phi(\vec{r}_{k})|^2 \Big[\oS_{z}\oI_{k,z}+\frac{1}{2}(\oS_{+}\oI_{k,-}+\oS_{-}\oI_{k,+})\Big]
\,,
\label{eq:HHI}
\end{equation}
where $|\phi(\vec{r}_{k})|^2$  is the probability to find the electron at the site of the $k$-th spin-carrying nucleus. The energy of the HI is given by a constant
\begin{equation}
A_{\HI}
=
g\cdot g_{N}\cdot C
\end{equation}
with $C>0$ being a material dependent energy scale. Since the HI interaction is strongly localized around the respective nucleus, the bare electron g-factor $g\approx2$ enters here\cite{Yafet1961,Coish2009}. Thus, the sign of the HI is determined by the sign of the nuclear g-factor $g_{N}$. This sign, however, plays an important role, since it determines the form of the ground state of the HI. 

If the coupling constant $A_{\HI}$ is positive (negative), the ground state of the bare HI will favor an anti-parallel (parallel) alignment of the electron spin with respect to all nuclear spins. Both ground states are twice degenerate, since a flip of all spins results in the same energy. Similarly, the two Zeeman terms also show the same two types of spin ordering depending on the signs of the respective g-factors $g^{*}$ and $g_{N}$. In contrast to the HI, these ground states are unique.  If, for instance, $g^{*}<0$ and $g_{N}>0$, the Zeeman terms would force both the electron spin and all nuclear spins to be parallel to the external magnetic field at zero temperature. Thus, when both the HI and the Zeeman interaction are present, there can arise an interesting competition of spin ordering with the electron spin being parallel or anti-parallel with respect to the nuclear spins. In particular, if the external magnetic field is sufficiently small, the HI is still strong enough to maintain the anti-parallel alignment of the electron spin with respect to the nuclear spins. If then additionally the signs of the g-factors are given by $g^{*}<0$ and $g_{N}>0$, the electron spin will be anti-parallel to the external magnetic field, whenever its Zeeman energy is below the total Zeeman energy of all nuclear spins. Starting from this particular ground state, we will show that a sudden flip of all spins can happen at a finite temperature $\Tzero>0$.

The article is organized as follows: In Sec. II.A, we will argue how the HI Hamiltonian can be simplified based on physical arguments. In Sec. II.B, we will then apply standard statistical physics to the Hamiltonian $\oH=\oH_{\ZE}^{S}+\oH_{\ZE}^{I}+\oH_{\HI}$, where we neglect the off-diagonal parts of the HI. Doing so, we calculate the thermal expectation value of the electron spin, whose properties are studied both analytically and numerically in Secs. II.C and II.D, respectively. After this mathematical analysis, we will then explain the physical mechanism being responsible for this spin flip in Sec. II.E. In Sec. III, we finally review our initial simplifications of the HI and discuss in which real systems our findings should be observable. In Apps. \ref{app:sec:diag} and \ref{app:sec:part_func}, we calculate the thermal expectation values for the full HI Hamiltonian including its off-diagonal part, i.e. the flip-flop terms. In App. \ref{app:sec:ground_state}, we analyze the behavior of the electron spin at zero temperature for this interaction.

\section{Thermal electron spin flip}
\label{sec:II}

\subsection{Simplified Hamiltonian}
\label{sec:sub:a}

As mentioned above, we will first introduce certain simplifications to the HI Hamiltonian, which allow for an analytical calculation of thermal expectation values:
\begin{enumerate}
\item We assume that all nuclear spins are spin one-half, where the number of nuclear spins is $K$.
\item We will use the so-called box-model, where the probabilities $|\phi(\vec{r}_{k})|^2=1/K$ are all the same. By this, we assume that every nucleus in the dot carries a spin and that the envelope function $\phi(\vec{r}_{k})$ of the electron does not change much inside the QD.
\end{enumerate}
With these two assumptions and $g^{*}<0$ as explained in the introduction, the total Hamiltonian is given by
\begin{align}
&
\oH_{\alpha}
=
\oH_{\ZE}^{S}+\oH_{\ZE}^{I}+\oH_{\HI}=
\nonumber\\
& 
-|g^{*}|\mu_{B}B_{z}\oS_{z} - g_{N}\mu_{N}B_{z}\oJ_{z}
\nonumber\\
& 
+ \frac{A_{\HI}}{K} [\oS_{z}\oJ_{z}+\frac{\alpha}{2}(\oS_{+}\oJ_{-}+\oS_{-}\oJ_{+})]
\,,
\label{eq:Hsitot}
\end{align}
where we have introduced the total nuclear spin $\vec{\oJ}=\sum_{k=1}^{K}\vec{\oI}_{k}$ for convenience. The usual raising and lowering operators $\oS_{\pm}=\oS_{x}\pm\I\oS_{y}$ and $\oJ_{\pm}=\oI_{x}\pm\I\oJ_{y}$ form the flip-flop terms. By means of the parameter $\alpha$ we distinguish between the full Hamiltonian ($\alpha=1$) and a simplified Hamiltonian ($\alpha=0$), which allows us to present the basic physics of the electron spin flip more easily.

Before we proceed with the calculation of thermal expectation values, we will introduce dimensionless units by measuring all energies in units of $\frac{A_{\HI}}{2K}$. The total Hamiltonian then reads
\begin{equation}
\oH_{\alpha}
=
-\sigma\oS_{z} - \nu\oJ_{z} + 2 \oS_{z}\oI_{k,z} + \alpha (\oS_{+}\oJ_{-}+\oS_{-}\oJ_{+})
\,,
\label{eq:Hsitot_dl}
\end{equation}
where the dimensionless parameters are given by $\sigma=|g^{*}|\mu_{B}B_{z}/\frac{A_{\HI}}{2K}$ and $\nu=g_{N}\mu_{N}B_{z}/\frac{A_{\HI}}{2K}$. These two quantities are obviously not independent of each other since both are proportional to the external magnetic field $B_{z}$. Thus we choose $\sigma=K\rho\nu$, where $\rho=|g^{*}|\mu_{B}B_{z}/(K\,g_{N}\mu_{N}B_{z})$ is the ratio of the Zeeman energies of the electron and all nuclear spins, respectively. This ratio can be also characterized by the critical number $\kappa=\rho K$, which is a constant for a given material.

In the following, we will first analyze the thermal expectation value of the electron spin for the simplified Hamiltonian $\oH_{0}$. Since this Hamiltonian is already diagonal in the basis of products states of the individual spin states, all calculations are much easier and, hence, the physics causing the spin flip becomes more apparent. However, it is not clear in the first place, if the neglected flip-flop terms give rise to quantum fluctuations, which destroy the electron spin flip. Thus, we also have calculated the thermal expectation value of the electron spin for the full Hamiltonian $\oH_{1}$ including the flip-flop terms. The details of these calculations are reported in Apps. \ref{app:sec:diag} to \ref{app:sec:ground_state}, while we will only present the respective results in the main text. Interestingly, many findings are unchanged with respect to the simpler case or restored in the limit of large system sizes, where the flip-flop terms are shown to be irrelevant. 

\subsection{Thermal expectation values}
\label{sec:sub:b}

\begin{figure}
\centering
\includegraphics[width=0.98\linewidth]{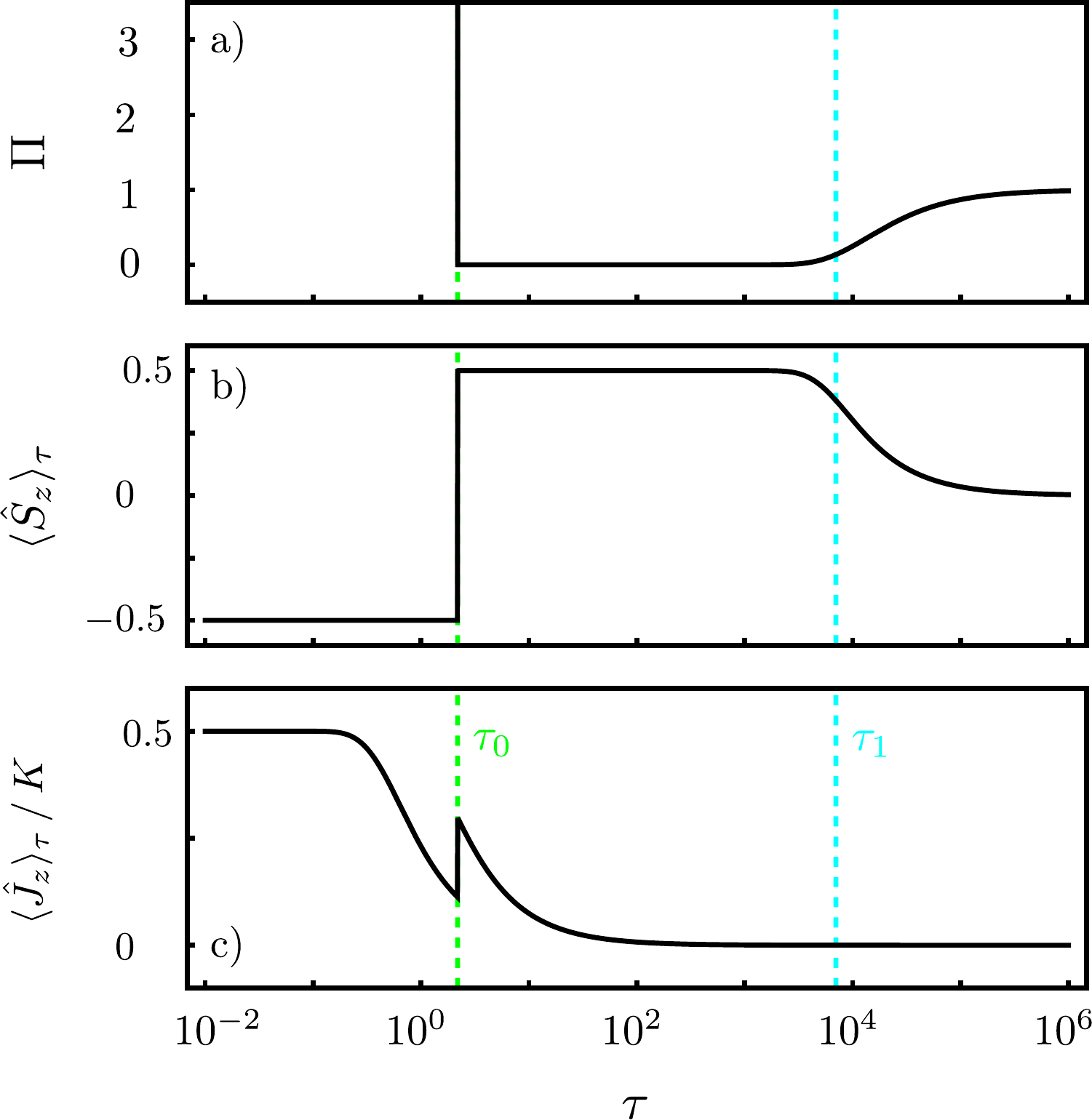}
\caption{(Color online) \textbf{a)}: $\Pi(\tau,\nu,\rho,K)$ as a function of temperature $\tau$ for $\nu=2$, $\rho=0.35$ and $K=10^{4}$. As we show in the text and in Fig.~\ref{fig:density_plot}, this choice of parameters fulfills the necessary conditions for the spin flip. At a temperature $\tauzero$, the function suddenly drops from very large values to zero. At temperatures $\tau\gtrsim\tauone$ the function surprisingly rises again and saturates at $\Pi=1$ for large temperatures. \textbf{b)}: The thermal expectation value of the electron spin $\av{\oS_{z}}_{\tau}$ exhibits a sudden flip at $\tauzero$. For temperatures above $\tauone$ the electron spin is thermally equilibrated. \textbf{c)}: The thermal expectation value of the total nuclear spin $\av{\oJ_{z}}_{\tau}=\frac{\tau}{2}\frac{\partial}{\partial \nu}\ln[Z]$ decreases before the electron spin flip. At $\tauzero$ it is suddenly increased again. For $\nu<1$ instead, the total nuclear spin would exhibit a flip similar to the electron spin.}
\label{fig:pi_sz}
\end{figure}

Without the flip-flop terms, the simplified Hamiltonian is already diagonal in the product basis $\ket{m_{S}}\otimes\bigotimes_{k=1}^{K}\ket{m_{k}}$, where $\ket{m_{S}}$ is an eigenstate of $\oS_{z}$ with $m_{S}\in\lbrace-1/2,1/2\rbrace$. Similarly, the state $\ket{m_{k}}$ is an eigenstate of the $k$-th nuclear spin operator $\oI_{k,z}$ with $m_{k}\in\lbrace-1/2,1/2\rbrace$. Thus, the partition function for this Hamiltonian is easily calculated in the product basis
\begin{align}
& Z
=
\tr{e^{-\oH_{0}/k_{B}T}}
=
\nonumber\\
&
\sum_{m_{S}}\sum_{\lbrace m_{k}\rbrace_{k=1}^{K}}e^{-\bra{m_{S},m_{K},m_{K-1},\dots}\oH_{0}\ket{m_{S},m_{K},m_{K-1},\dots}/k_{B}T}
\,,
\label{eq:Z}
\end{align}
where the diagonal matrix elements are given by
\begin{align}
&-\bra{m_{S},m_{K},m_{K-1},\dots}\oH_{0}\ket{m_{S},m_{K},m_{K-1},\dots}/k_{B}T
\nonumber\\
& =\frac{2}{\tau}[\rho\nu K\, m_{S}+\sum_{k=1}^{K}\;(\nu\,m_{k}-2\, m_{S}m_{k})]
\end{align}
with the dimensionless temperature $\tau=k_{B}T/\frac{A_{\HI}}{2K}$. Exploiting the fact that sums in the exponential functions factorize finally yields
\begin{align}
Z
& =
\sum_{m_{S}}\prod_{k=1}^{K}\sum_{m_{k}} e_{m_{S},m_{k}}
\nonumber\\
& =
(e_{\frac{1}{2},\frac{1}{2}} + e_{\frac{1}{2},-\frac{1}{2}})^{K} + (e_{-\frac{1}{2},\frac{1}{2}} + e_{-\frac{1}{2},-\frac{1}{2}})^{K}
\,,
\label{eq:Z_final}
\end{align}
where
\begin{equation}
e_{m_{S},m_{k}}
=
\exp[ \frac{2}{\tau}\,( \rho\nu\,m_{S} + \nu\,m_{k} -2\,m_{S}m_{k} )]
\,.
\label{eq:e_func}
\end{equation}
With the partition function at hand, the calculation of the thermal expectation value of the electron spin is readily obtained
\begin{align}
\av{\oS_{z}}_{\tau}
& =
\frac{\tau}{2}\frac{\partial}{\partial \sigma}\ln[Z]
=
\frac{\tau}{2K\nu}\frac{\partial}{\partial \rho}\ln[Z]
\nonumber\\
& =
\frac{1}{2}\Big\lbrace\frac{1}{1-\Pi}-\frac{1}{1-\Pi^{-1}}\Big\rbrace
\,,
\label{eq:avSz}
\end{align}
where the function
\begin{align}
\Pi
& =
\Pi(\tau,\nu,\rho,K)
=
\Bigg[\frac{e_{-\frac{1}{2},\frac{1}{2}} + e_{-\frac{1}{2},-\frac{1}{2}}}{e_{\frac{1}{2},\frac{1}{2}} + e_{\frac{1}{2},-\frac{1}{2}}}\Bigg]^{K}
\label{eq:Pi}
\end{align}
controls the behavior of $\av{\oS_{z}}_{\tau}$. If $\Pi\to\infty$, we find $\av{\oS_{z}}_{\tau}\to-1/2$, $\Pi=0$ results in $\av{\oS_{z}}_{\tau}=1/2$ and, finally, $\Pi=1$ yields $\av{\oS_{z}}_{\tau}=0$. As it turns out, the electron spin has to go through exactly these steps for the thermal spin flip to occur as illustrated in Fig.~\ref{fig:pi_sz}.
In the following, we will first explore the parameter space of $\Pi(\tau,\nu,\rho,K)$ to find mathematically the conditions necessary for the spin flip to occur. Afterwards we calculate at which temperatures $\Pi$ and, consequently, $\av{\oS_{z}}_{\tau}$ undergo their characteristic changes. To this end, we will analyze $\Pi$ analytically and compare the findings with numerical calculations of $\av{\oS_{z}}_{\tau}$. Finally, we will interpret these results in order to understand when and why this effect appears in a physical system.

\subsection{Analytical analysis of the spin flip}
\label{sec:sub:c}

By inserting the definition of the exponential functions in Eq.~(\ref{eq:e_func}) into Eq.~(\ref{eq:Pi}) and rearranging factors, we find
\begin{align}
&
\pi^{K}
\equiv
\Pi(\tau,\nu,\rho,K)=
\nonumber\\
& 
\Bigg[\exp[-\frac{2\rho\nu}{\tau}]\frac{\exp[\frac{1}{\tau}\lbrace-\nu-1\rbrace] + \exp[\frac{1}{\tau}\lbrace\nu+1\rbrace]}{\exp[\frac{1}{\tau}\lbrace-\nu+1\rbrace]+\exp[\frac{1}{\tau}\lbrace\nu-1\rbrace]}\Bigg]^{K}
\,.
\label{eq:Pi_rear}
\end{align}
By further rearrangements of factors in Eq.~(\ref{eq:Pi_rear}), we identify that $\rho<1$ and $0<\nu<\rho^{-1}$ are necessary conditions for $\Pi$ to diverge at $\tau\to0$ and, consequently, for the spin flip to occur. Within this parameter regime, we want to identify the temperature $\tauzero$, at which $\Pi$ drops from infinity to zero, and the temperature $\tauone$ at which $\Pi$ rises to $1$ as indicated in Fig.~\ref{fig:pi_sz}. 

The latter temperature can be readily read off from Eq.~(\ref{eq:Pi_rear}), since $\Pi=1$ for all temperatures $\tau$ well above
\begin{equation}
\tauone
=
2\rho\nu K
\,.
\label{eq:tau_1}
\end{equation}

At a specific temperature $\tau$ the function $\pi$ defined in Eq.~(\ref{eq:Pi_rear}) changes from $\pi>1$ to $\pi<1$. Since one has to take it to the power of $K\gg1$, this marks the temperature, at which the sudden drop from $\Pi\gg1$ to $\Pi\ll1$, and, hence, the spin flip occurs. Thus, the transcendental equation defining $\tauzero$ reads
\begin{equation}
\rho
=
-\frac{\tauzero}{2\nu}\ln\Bigg[\frac{\exp[\frac{1}{\tauzero}\lbrace-\nu-1\rbrace] + \exp[\frac{1}{\tauzero}\lbrace\nu+1\rbrace]}{\exp[\frac{1}{\tauzero}\lbrace-\nu+1\rbrace]+\exp[\frac{1}{\tauzero}\lbrace\nu-1\rbrace]}\Bigg]
\,.
\label{eq:tau_infty}
\end{equation}
This equation can be expanded in powers of $\frac{1}{\tauzero}\ll1$,
\begin{equation}
\rho
\approx
\frac{1}{\tauzero}+\mathrm{O}(\frac{1}{\tauzero^{3}})
\,.
\end{equation}
As a consequence, the temperature $\tauzero\approx\rho^{-1}$ is independent of $\nu$ for $\rho\ll1$. Since $\rho=\kappa/K$ is a constant for a given QD, this is a rather intriguing result. This constant being $\rho\ll1$ corresponds to a situation, where the total nuclear Zeeman energy is much larger than the electron Zeeman energy. 

Before we give a detailed physical interpretation of our results, let us compare these analytical results for the simplified Hamiltonian $\oH_{0}$ with i) a numerical analysis of Eqs.~(\ref{eq:avSz}) and (\ref{eq:Pi}), respectively, and ii) the behavior of the electron spin for the full Hamiltonian $\oH_{1}$ calculated in Apps. \ref{app:sec:diag} to \ref{app:sec:ground_state}.

\subsection{Numerical analysis of the spin flip} 
\label{sec:sub:d}

\begin{figure*}
\centering
\includegraphics[width=\textwidth]{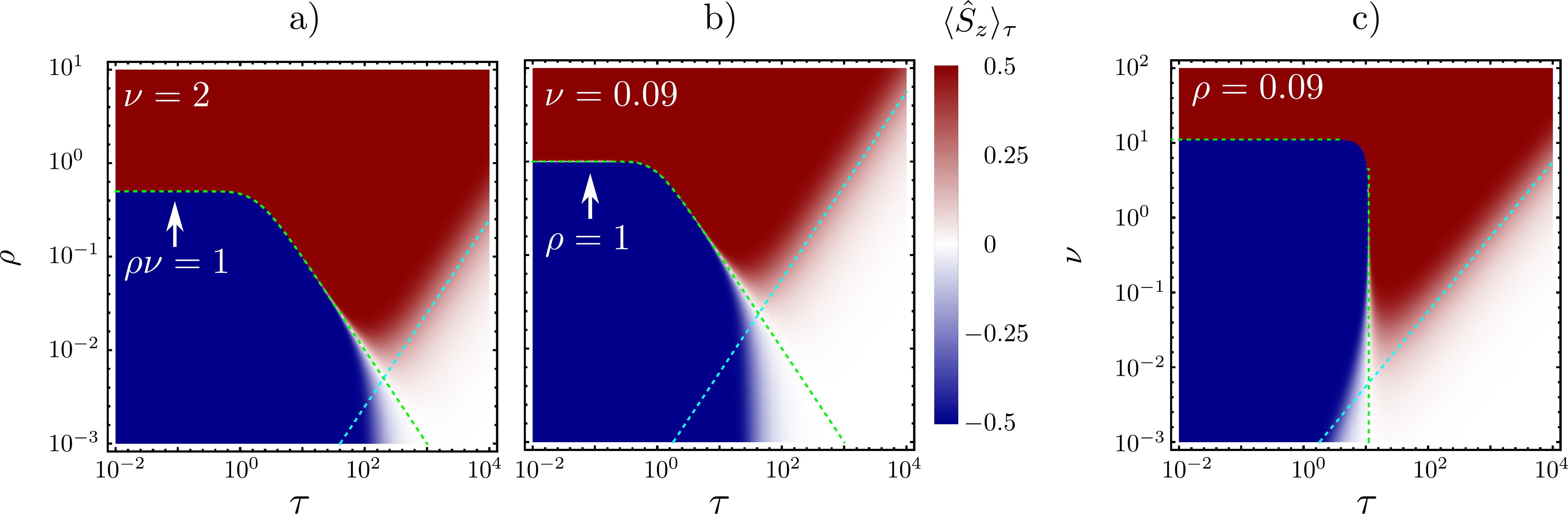}
\caption{(Color online) \textbf{a)}: The thermal expectation value of the electron spin as a function of temperature $\tau$ and $\rho$ for $\nu=2$. The green dashed line shows the defining Eq.~(\ref{eq:tau_infty}) of $\tauzero$. For small $\rho$ one finds $\tauzero=\rho^{-1}$ in agreement with the plot.  The light blue line is given by $\tauone$ in Eq.~(\ref{eq:tau_1}). For $\rho\nu\geq1$ the spin flip is absent. For temperatures above $\tauone$, the electron is thermally equilibrated and, hence, $\av{\oS_{z}}_{\tau}=0$. \textbf{b)}: The thermal expectation value of the electron spin as a function of temperature $\tau$ and $\rho$ for $\nu=0.09$. Clearly, the spin flip is absent for $\rho\geq1$. \textbf{c)}: The thermal expectation value of the electron spin as a function of temperature $\tau$ and $\nu$ for $\rho=0.09$. The light blue line is given by $\tauone$ in Eq.~(\ref{eq:tau_1}). The horizontal green line corresponds to $\nu=\rho^{-1}$ and the vertical green line to $\tau=\rho^{-1}$. For $\rho\nu\geq1$ the spin 
flip is absent. For temperatures above $\tauone$, the electron is again thermally equilibrated.}
\label{fig:density_plot}
\end{figure*}

In order to verify our analytical results for $\oH_{0}$, we show density plots of $\av{\oS_{z}}_{\tau}=\av{\oS_{z}}_{\tau}(\tau,\nu,\rho,K)$, where we choose the number of nuclear spins to be $K=10^4$. With $K$ being fixed, $\tau$, $\rho$, and $\nu$ remain as parameters. In Fig.~\ref{fig:density_plot}~a), we show $\av{\oS_{z}}_{\tau}$ as a function of $\tau$ and $\rho$ with $\nu=2$. As we will show below, for this choice of $\nu$, the electron Zeeman energy competes with the total HI energy. If $\rho\nu>1$ the Zeeman energy exceeds the HI energy and the electron spin is up for all $\tau$, which is clearly shown in Fig.~\ref{fig:density_plot}~a). Additionally, we plotted Eqs.~(\ref{eq:tau_1}) and (\ref{eq:tau_infty}) in order to demonstrate the behavior of $\tauone$ and $\tauzero$, respectively. Both analytical results show a remarkable agreement with the numerical findings.  In Fig.~\ref{fig:density_plot}~b), we show the same plot for $\nu=0.09$. For this choice of $\nu$, the electron Zeeman energy is always smaller than the total HI energy. However, if $\rho>1$, the electron Zeeman energy is larger than the total nuclear Zeeman energy and, consequently, the electron spin flip is up for all $\tau$ as can be seen from Fig.~\ref{fig:density_plot}~b).

As indicated above, in a real system $\rho$ is rather a fixed parameter than a real variable. Hence, we also calculated $\av{\oS_{z}}_{\tau}$ as a function of temperature $\tau$ and the Zeeman energy of a single nucleus $\nu$, which is proportional to the magnetic field $B_{z}$. The result is shown in Fig.~\ref{fig:density_plot}~c), where we chose $\rho=0.09$. In this figure, the behavior of $\tauzero=\rho^{-1}$ is most prominent. Moreover, it is obvious that $\rho\nu<1$ is indeed a necessary condition for the spin flip.

Finally, we have confirm that the influence of the flip-flop terms, which are present in the full Hamiltonian $\oH_{1}$ does not destroy the electron spin flip. As we show in Apps. \ref{app:sec:diag} and \ref{app:sec:part_func}, the thermal expectation value of the electron spin can be exactly solved for the box-model.\cite{Zhang2006,Coish2007a,Bortz2007,Erbe2010} By investigating the temperature dependence of $\av{\oS_{z}}_{\tau}$ for up to $K=60$ nuclear spins, we see that both temperature scales $\tauzero$ and $\tauone$ are unchanged. For small numbers of nuclear spins, however, we find that the additional interaction alters the behavior of $\av{\oS_{z}}_{\tau}$. To be more specific, the minimum of the thermal expectation value $\av{\oS_{z}}_{0}$ at zero temperature is larger than $-1/2$ for few nuclear spins as can be seen from Fig. \ref{app:fig:Sz_0_K}. For larger system sizes, the original value of $\av{\oS_{z}}_{0}=-1/2$ seems to be restored. Identifying the ground state of the system in App. \ref{app:sec:ground_state}, we are indeed able to show that $\av{\oS_{z}}_{0}=-1/2$ is exactly reached in the limit of large $K$. Moreover, the maximum of the electron expectation value at approximately $\tau=2/\rho$ is altered by the flip-flop terms as is obvious from Fig. \ref{app:fig:Sz_rho_K}. Again, their effect is most pronounced for small $K$, while the results for $\oH_{0}$ are reproduced for large system sizes. Thus, the quantum fluctuations do not destroy the spin flip. Physically, it seems very likely, that the nuclear Zeeman energy additionally stabilizes the spin system against the flip-flop terms. In the limit of large system sizes, the physics of the simplified Hamiltonian is restored, which can be understood by analyzing how different states are affected by the flip-flop terms. The states being most efficient for this interaction are of the form $\ket{-1/2,1/2,\dots,1/2}$ and vice versa, since the electron spin can flip with every nuclear spin. The states $\ket{-1/2,1/2,-1/2,1/2,-1/2,\dots}$ (and all permutations of the nuclear spins) are the less affected ones. For large system sizes, however, the statistical weight of the latter states is much higher than for the former states, explaining the vanishing influence of the flip-flop terms for increasing $K$.

\subsection{Physical interpretation of the results}
\label{sec:sub:e}

\begin{figure}
\centering
\includegraphics[width=0.98\linewidth]{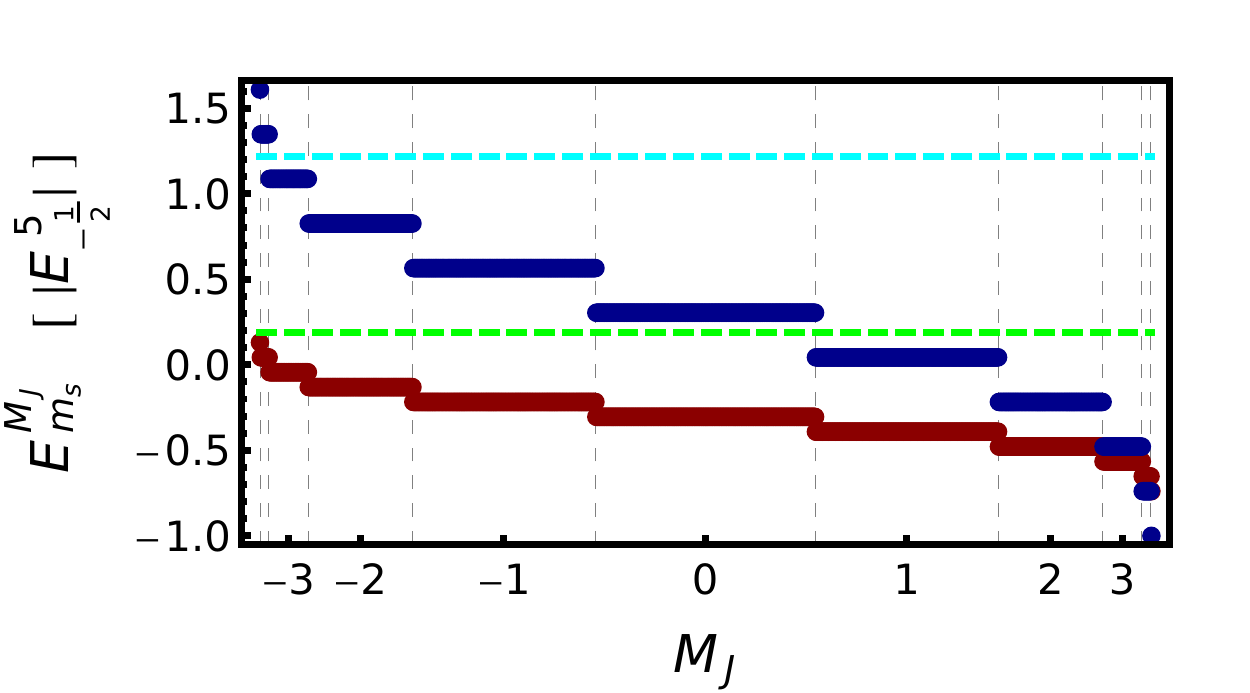}
\caption{(Color online) Energies of the Hamiltonian as a function of the total nuclear angular momentum $M_{J}=\sum_{k}m_{k}$ for electron spin up ($m_{S}=1/2$, red) and down ($m_{S}=-1/2$, blue). The other parameters are $\nu=2$, $\rho=0.35$, and $K=10$. The width of a column is proportional to the degeneracy of the respective energy level given by $N_{M_{J}}^{K}=\binom{K}{K/2-M_{J}}$. The green and light blue line show the temperatures $\tauzero$ and $\tauone$, respectively.}
\label{fig:qen_plot_Rho__0_35_Nu_2_0_Knuc__10}
\end{figure}

So far, we have mathematically clarified for which parameters one finds the spin flip. In the following, we want to explain why this spin flip occurs and how the conditions found above can be interpreted physically. Therefore, we will have a closer look on the energies of the Hamiltonian $\oH_{0}$ given in Eq.~(\ref{eq:Hsitot_dl}). Since the Hamiltonian is invariant under the exchange of two nuclear spins, only the total nuclear angular momentum $M_{J}=\sum_{k}m_{k}$ is relevant resulting in $N_{M_{J}}^{K}=\binom{K}{\frac{K}{2}-M_{J}}$ equal energies of size
\begin{equation}
E_{m_{S}}^{M_{J}}
=
-K\rho\nu m_{S} - M_{J}\nu + 2m_{S}M_{J}
\,.
\label{eq:energy}
\end{equation}
A plot of $E_{1/2}^{M_{J}}$ and $E_{-1/2}^{M_{J}}$ is presented in Fig.~\ref{fig:qen_plot_Rho__0_35_Nu_2_0_Knuc__10}, where a small system size of $K=10$ nuclear spins has been chosen for practical reasons. Although the spin flip is not perfect for such a small number of nuclear spins as is obvious from Fig. \ref{app:fig:Sz_rho_K}, the physics of the spin flip still becomes clear. The ground state is given by $m_{S}=-1/2$ and $M_{J}=K/2$, whose energy eigenvalue $E_{-1/2}^{K/2}$ is not degenerate. If one follows the energies $E_{-1/2}^{M_{J}}$ and $E_{1/2}^{M_{J}}$ starting from $M_{J}=K/2$, one finds that $E_{-1/2}^{M_{J}}$ increases much faster than $E_{1/2}^{M_{J}}$, which is also obvious from Eq.~(\ref{eq:energy}). Since the degeneracy of the corresponding energy levels $N_{M_{J}}^{K}$ is strongly increasing, many states with $m_{S}=1/2$ become thermally available for finite temperatures. Once the temperature reaches $\tauzero$, there is a strong imbalance between the number of states with $m_{S}=-1/2$ and the number of states with $m_{S}=1/2$, which finally causes the sudden spin flip. If the temperature is further increased above $\tauone$ almost all states are reached and, hence, one finds $\av{\oS_{z}}_{\tau}=0$. 

Finally, we would like to interpret the mathematical conditions on the parameters physically. Let us start with the constraints on $\rho$ and $\rho\nu$. Since the  $\rho=|g^{*}|\mu_{B}B_{z}/(Kg_{N}\mu_{N}B_{z})$ is given by the ratio of the electron Zeeman and the total nuclear Zeeman energy, the ratio being $\rho<1$ implies that the Zeeman energy of all nuclear spins exceeds the electron Zeeman energy. Similarly the product $\rho\nu=2|g^{*}|\mu_{B}B_{z}/A_{\HI}$ tells us that the Zeeman energy of the electron has to be smaller than the total HI energy. This imposes an upper bound on the external magnetic field
\begin{equation}
B_{z}
<
B_{\mymax}
=
\frac{A_{\HI}}{2|g^{*}|\mu_{B}}
\,.
\end{equation}
But the magnetic field has additionally to be large enough in order to separate the two temperatures $\tauzero$ and $\tauone$. For small magnetic fields, the former temperature is given by $\tauzero=\rho^{-1}$, which corresponds to an absolute temperature
\begin{equation}
\Tzero=k_{B}^{-1}\frac{A_{\HI}}{2}\;\frac{g_{N}\mu_{N}}{|g^{*}|\mu_{B}}
\,.
\label{eq:T_infty}
\end{equation}
The latter temperature $\tauone$ corresponds to the Zeeman splitting of the electron since
\begin{equation}
\Tone=k_{B}^{-1}|g^{*}|\mu_{B}B_{z}
\,.
\label{eq:T_1}
\end{equation}
For $\Tzero<\Tone$ the spin flip is present, which is the case if the magnetic field obeys
\begin{equation}
B_{z}
>
B_{\mymin}
=
\frac{g_{N}\mu_{N}}{|g^{*}|\mu_{B}}\;\frac{A_{\HI}}{2|g^{*}|\mu_{B}}
=
\frac{1}{\kappa}B_{\mymax}
\,.
\end{equation}

\section{Discussion and outlook}
\label{sec:III}

\begin{table}
\renewcommand{\arraystretch}{1.5}
\begin{ruledtabular}
\begin{tabular}{l|*{8}{>{$}c<{$}}}
		& g^{*}		& g_{N}		& \kappa		& A_{\HI}		& B_{\mymax}			& I_{\mymin}		& \Tzero 		& \Tone			\\
Material	& 		& 		& [10^{3}]			& [\mu\mathrm{eV}]	& [\mathrm{mT}]		& 					& [\mathrm{mK}]		& [\mathrm{mK}]		\\
\\\hline\\
GaAs		& -0.4		& 1.8		& 0.44			& 84			& 1700			& \frac{3}{2}				& 5.6	& 49	\\
CdTe		& -1.8		& -6.7		& 4.9			& -34			& 170			& \frac{1}{2}				& 0.04	& 20			\\
GaInAs\footnote{$\mathrm{Ga}_{0.47}\mathrm{In}_{0.53}\mathrm{As}$}
		& -4.4		& 3.1		& 2.6			& 93			& 180	& \frac{3}{2}		& 1.0	& 54	\\
InAs		& -15		& 3.5		& 7.9			& 98			& 57	& \frac{3}{2}						& 0.36	& 57	\\
\end{tabular}
\end{ruledtabular}
\caption{Relevant materials and their parameters. The values of $g^{*}$ are taken from Ref~\onlinecite{Winkler2003}. The magnetic moment $g_{N}$ and the HI constant $A_{\HI}$ are averaged by $g_{N}=\sum_{i}n_{i}g_{N}^{i}$ and $A_{\HI}=\sum_{i}n_{i}A_{\HI}^{i}$, where $n_{i}$ is the natural abundance of isotope $i$. The values are taken from Refs.~\onlinecite{Schliemann2003,Testelin2008,Coish2009}. The temperatures $\Tone$ and $\Tzero$ are calculated at $B_{z}=0.1\,B_{\mymax}$, at which $\Tzero=k_{B}^{-1}\cdot[4I(I+1)/3]\cdot\kappa^{-1}\frac{A_{\HI}}{2}$ is valid. By this choice, we also take into account that larger nuclear spin quantum numbers $I$ increase $\Tzero$. For simplicity, we took the smallest quantum number $I_{\mymin}$ of different isotopes present in the dot.}
\label{tab:materials}
\end{table}

All of the above results were obtained applying certain approximations, which are in general not fulfilled by real systems. In the following, we will discuss all approximations one by one and analyze, how a more realistic model would change our results.

First off all, we assumed that all nuclear spins are of the same species. This is a commonly made approximation\cite{Coish2009}, where root-mean-square averages of the Zeeman and HI coupling constants can be used to mimic the situation of only one spin species being present. As long as the individual constants are not too different and if the sign of the nuclear magnetic moment is the same for all nuclear spins, we do not expect qualitative changes of our findings. Additionally, we have chosen all nuclear spins to be one-half. If one allows for larger nuclear spin quantum numbers $I=1/2,3/2,\dots$, the model can still be solved analytically. Since the thermal relaxation of the electron spin does only depend on its Zeeman energy, the temperature $\tauone$ is unchanged. In contrast to this, the spin flip temperature $\tauzero=\rho^{-1}4\,I(I+1)/3$ increases by an $I$-dependent factor. Finally, we also implicitly assumed that all nuclei carry a spin. Yet, this is not the case for all materials. If $K$ out of $N$ nuclei carry a spin, the probability to find the electron at the site of a nuclear spin is $|\phi(\vec{r}_{k})|^2=N^{-1}=K^{-1}\cdot(K/N)\equiv K^{-1}\cdot n_{I}$, where $0\leq n_{I}\leq1$ is the abundance of spin carrying nuclei.\cite{Fischer2009a} Hence, our results still hold if the HI constant $A_{\HI}$ is replaced by $n_{I}\,A_{\HI}$.

Besides approximations concerning the nuclear spins, we also simplified the physics of the electron spin by using the box model for its envelope function: $|\phi(\vec{r}_{k})|^2=K^{-1}$. In reality, the probability to find the electron, should strongly decrease with the distance from the center of the QD, which is often described by a Gaussian envelope function $|\phi(\vec{r}_{k})|^2\propto K^{-1}\exp[-r_{k}/R]$. Our results should be modified in this case by two aspects: Nuclear spins with $|\phi(\vec{r}_{k})|^2\ll K^{-1}$ couple only very weakly to the electron and can, thus, be neglected. Effectively, this reduces the number of involved nuclear spins from $K$ to $K_{\mathrm{eff}}<K$. For nuclear spins in the center, one finds $\mathcal{O}(|\phi(\vec{r}_{k})|^2)=K^{-1}$. As a consequence, the non-uniform HI will (slightly) lift the degeneracy of the energies in Eq.~(\ref{eq:energy}) and Fig.~\ref{fig:qen_plot_Rho__0_35_Nu_2_0_Knuc__10}, but it will not change the energy spectrum in principle. Therefore, the main physical mechanism stays the same and our results still hold. 

Finally, we have investigated the behavior of the electron spin expectation value for the full Hamiltonian including the HI flip-flop terms in Apps. \ref{app:sec:diag} to \ref{app:sec:ground_state}. Doing so, we have confirmed, that the relevant temperature scales are essentially the same and that the physics of the simplified Hamiltonian is reproduced in the limit of a large number of nuclear spins.

Having convinced ourselves, that the results obtained within the simplified model should be reasonable for real systems, we finally want to give several examples, where we expect the spin flip to occur. The most severe constraint is the negative sign of $g^{*}$. As can be seen from Tab. \ref{tab:materials}, this is realized, for instance, in III-V heterostructures, where the electron experiences a strong spin-orbit interaction. Most promising among the considered materials is GaAs, since its spin flip temperature is on the order of mK. The other materials having a smaller $\Tzero$ suffer mostly from a large $g^{*}$ factor. Beside a small $g^{*}$ factor, potential materials would also benefit from a strong HI and from heavy nuclei with large $g_{N}$ factors and large spin quantum numbers $I$. Among them, also systems with a negative $g_{N}$ such as CdTe can be considered, since this sign changes both the nuclear Zeeman coupling and the sign of the HI. Redefining the nuclear spin operator by $\oI_{z} \rightarrow-\oI_{z}$ then yields the same results.

Finally, we will briefly discuss the nature of the spin flip. If one leaves the equilibrium state while heating up the system (non-adiabatically), it will take some time for the system to reach the equilibrium at its new temperature. Especially for crossing $\Tzero$, the system is not only forced to flip the electron spin, but also approximately up to $\kappa\gg1$ nuclear spins. Thus, depending, for instance, on the microscopic details of the coupling of the spins to one or several baths, the time needed to equilibrate could be comparably long.

\section*{Acknowledgments}

The authors thank A. P\'{a}lyi for useful discussions and comments. We acknowledge financial support by the DFG (SSP 1449 and grant TR950/8-1).

\appendix

\section{Diagonalization of the full Hamiltonian}
\label{app:sec:diag}

In this section, we show how the full Hamiltonian $H_{\alpha}\equiv\oH(\sigma,\nu,\alpha)$ in Eq.~(\ref{eq:Hsitot_dl}) can be diagonalized.\cite{Zhang2006,Coish2007a,Bortz2007,Erbe2010} Like the simplified Hamiltonian $H_{0}$, it is invariant under an exchange of two nuclear spins. Moreover, a total flip of all spins ($\oF: \oF\ket{m_{S},m_{K},\dots,m_{1}}=\ket{-m_{S},-m_{K},\dots,-m_{1}}$) yields $\oF\oH(\sigma,\nu,\alpha)\oF^{\dagger}=\oH(-\sigma,-\nu,\alpha)$. Finally, the Hamiltonian commutes with the $z$-component of the total spin $\oS_{z}+\oJ_{z}$.\cite{Zhang2006,Coish2007a} Due to this property, it is convenient to represent the full Hamiltonian in the basis of product states between the electron spin and the total nuclear spin $\ket{m_{S}}\otimes\ket{J,M_{J},\lbrace q_{i}\rbrace}=\ket{m_{S},J,M_{J},\lbrace q_{i}\rbrace}$, in which it has a simple block-diagonal structure. The additional quantum numbers $\lbrace q_{i}\rbrace$ are related to the corresponding Clebsch-Gordon coefficients\cite{Bortz2007,Erbe2010}. These quantum numbers $q_{i}=q_{i}(\lbrace I_{k}\rbrace_{k=1}^{K})$ depend on the quantum numbers $I_{k}$ of the original states $\ket{I_{k},m_{k}}$. Since the Hamiltonian is degenerate in these quantum numbers, we will use the shorthand notation $\ket{J,M_{J}}$ for the nuclear spin state whenever it is appropriate. Ordering these product states $\ket{m_{S},J,M_{J}}$ properly, this yields a block-diagonal representation of the Hamiltonian, where the blocks are of dimension $1$ or $2$.

In this basis, all diagonal entries of the Hamiltonian are formed by the energies
\begin{align}
E_{1}(\sigma,\nu,M_{J},m_{S})
& =
\bra{m_{S},J,M_{J}}H(\sigma,\nu,\alpha)\ket{m_{S},J,M_{J}}
\nonumber\\
& =
-\sigma m_{S} -\nu M_{J} + 2 m_{S}M_{J}
\,,
\label{app:eq:E1}
\end{align}
where $m_{S}=1/2,-1/2$. The states $\ket{1/2,J,J}$ and $\ket{-1/2,J,-J}$ are already eigenstates of the Hamiltonian, and, hence the corresponding energies constitute the one-dimensional blocks. The off-diagonal parts of the two-dimensional blocks are given by
\begin{align}
F(J,M_{J}) & = \bra{-\frac{1}{2},J,M_{J}}H(\sigma,\nu,\alpha)\ket{\frac{1}{2},J,M_{J}-1}
\nonumber\\
& =
\sqrt{J(J+1)-M_{J}(M_{J}-1)}
\end{align}
Thus, the eigenenergies and eigenstates of the two-dimensional blocks are obtained by diagonalizing the $2\times2$ matrices
\begin{align}
& H_{2}(\sigma,\nu,\alpha,J,M_{J})
=
\nonumber\\
& \left(
\renewcommand*{\arraystretch}{1.5}
\begin{array}{cc}
E_{1}(\sigma,\nu,M_{J},-\frac{1}{2}) & \alpha F(J,M_{J}) \\
\alpha F(J,M_{J}) & E_{1}(\sigma,\nu,M_{J}-1,\frac{1}{2}) 
\end{array}
\right)
\,.
\label{app:eq:H2}
\end{align}
The eigenenergies of the $2\times2$ matrix in Eq.~(\ref{app:eq:H2}) are, then, given by
\begin{align}
& E_{2,\pm}(\sigma,\nu,\alpha,J,M_{J})=
\nonumber\\
& \frac{1}{2}[ E_{1}(\sigma,\nu,M_{J},-\frac{1}{2})+E_{1}(\sigma,\nu,M_{J}-1,\frac{1}{2})\pm
\nonumber\\
& \lbrace[E_{1}(\sigma,\nu,M_{J},-\frac{1}{2})-E_{1}(\sigma,\nu,M_{J}-1,\frac{1}{2})]^{2}
\nonumber\\
& +4\alpha^{2}F(J,M_{J})^{2}\rbrace^{\frac{1}{2}}
\label{app:eq:E2}
\end{align}

\section{Calculation of the partition function}
\label{app:sec:part_func}

\begin{figure}
\centering
\includegraphics[width=0.98\linewidth]{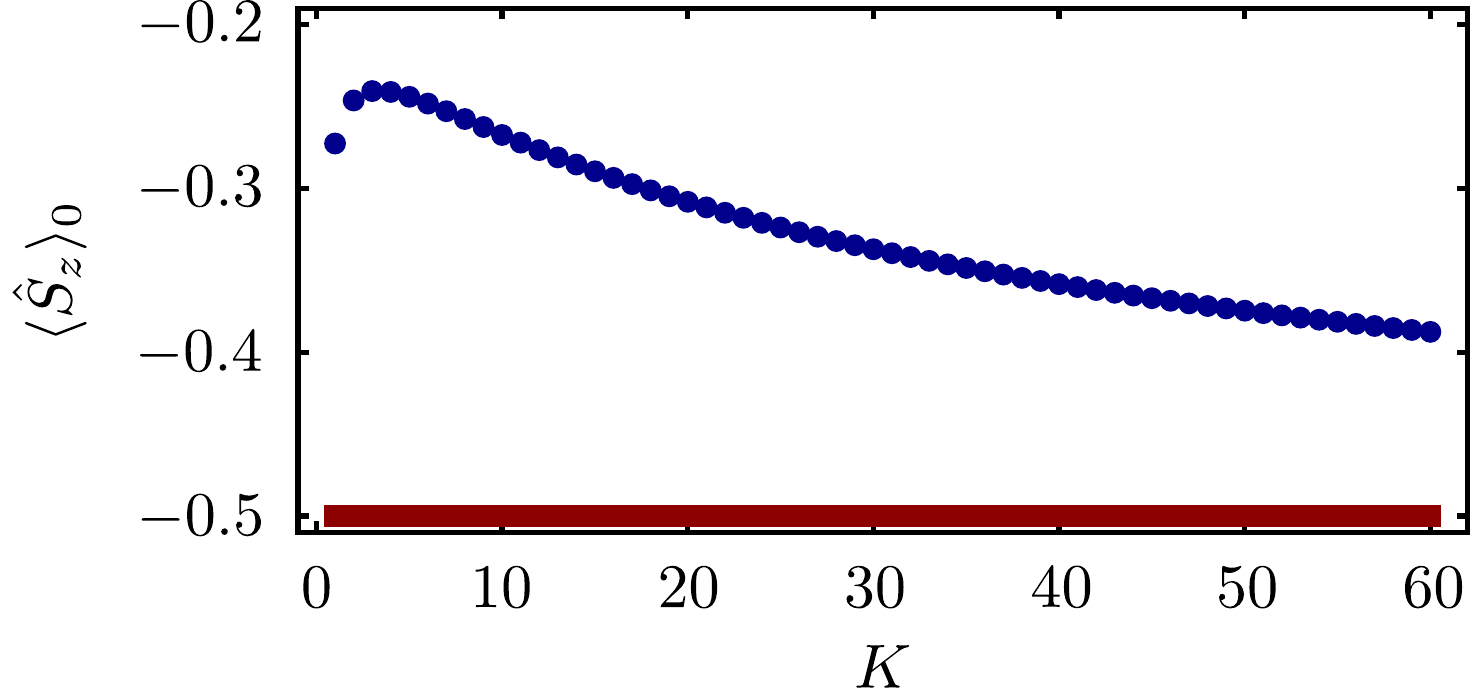}
\caption{(Color online) Dependence of the thermal expectation value of the electron spin $\av{\oS_{z}}_{\tau}$ at $\tau=0$ on the number of nuclear spin with (blue circles) and without (red squares) the flip-flop terms. The remaining parameters are $\rho=0.35$, $\nu=2$, and $\sigma=K\rho\nu$.}
\label{app:fig:Sz_0_K}
\end{figure}

In this section, we want to calculate the partition function $Z=\tr{e^{-\beta \oH_{\alpha}}}$ of the full Hamiltonian $\oH_{\alpha}\equiv\oH(\sigma,\nu,\alpha)$ in Eq.~(\ref{eq:Hsitot_dl}). Since the Hamiltonian is invariant under an exchange of two nuclear spins, the trace can be written as
\begin{align}
Z
=
\sum_{m_{S}} \sum_{M_{J}=-\frac{K}{2}}^{\frac{K}{2}}N_{M_{J}}^{K}\bra{m_{S},\Psi_{M_{J}}}e^{-\beta \oH_{\alpha}}\ket{m_{S},\Psi_{M_{J}}}
\,,
\end{align}
where the nuclear spin state is given by
\begin{equation}
\ket{\Psi_{M_{J}}}
=
\ket{\underbrace{-\frac{1}{2},\dots,-\frac{1}{2}}_{\frac{K}{2}-M_{J}},\underbrace{\frac{1}{2},\dots,\frac{1}{2}}_{\frac{K}{2}+M_{J}}}
\end{equation}
and $N_{M_{J}}^{K}=\binom{K}{\frac{K}{2}-M_{J}}$. In order to benefit from previous results\cite{Bortz2007,Erbe2010}, which have been obtained for $M_{J}\geq0$, we additionally use the effect of a total flip of all spins $\oF\ket{m_{S},\Psi_{M_{J}}}=\ket{-m_{S},\Psi_{-M_{J}}}$ and $\oF\oH(\sigma,\nu,\alpha)\oF^{\dagger}=\oH(-\sigma,-\nu,\alpha)$.
Thus, the partition function can be expressed by
\begin{align}
&Z
=
\nonumber\\
& \sum_{m_{S}}[
\sum_{M_{J}=\frac{\mymod{K}{2}}{2}}^{\frac{K}{2}}N_{M_{J}}^{K}\bra{m_{S},\Psi_{M_{J}}}e^{-\beta \oH(\sigma,\nu,\alpha)}\ket{m_{S},\Psi_{M_{J}}}+
\nonumber\\
&
\sum_{M_{J}=\frac{1-\mymod{K}{2}}{2}}^{\frac{K}{2}}N_{M_{J}}^{K}\bra{m_{S},\Psi_{M_{J}}}e^{-\beta \oH(-\sigma,-\nu,\alpha)}\ket{m_{S},\Psi_{M_{J}}}]
\,.
\end{align}
Finally, we replace the states $\ket{m_{S},\Psi_{M_{J}}}$ by the states $\ket{m_{S},J,M_{J}}$. For spin one-half particles, this replacement\cite{Bortz2007,Erbe2010} is given by
\begin{equation}
\ket{m_{S},\Psi_{M_{J}}}
=
\sum_{k=0}^{\frac{K}{2}-M_{J}}\sum_{\lbrace q_{i}\rbrace} c_{k}^{\lbrace q_{i}\rbrace}\ket{m_{S},\frac{K}{2}-k,M_{J},\lbrace q_{i}\rbrace}
\,
\end{equation}
where the coefficients $c_{k}^{\lbrace q_{i}\rbrace}$ obey the following relation
\begin{equation}
\sum_{\lbrace q_{i}\rbrace} \vert c_{k}^{\lbrace q_{i}\rbrace}\vert^{2}
=
\frac{(\frac{K}{2}-M_{J})!(\frac{K}{2}+M_{J})!(K-2k+1)}{(K-k)!k!(K-k+1)}
\equiv d_{k}
\end{equation}
The states $\ket{1/2,M_{J},M_{J},\lbrace q_{i}\rbrace}$ are already eigenstates of the Hamiltonian. Moreover, the state $\ket{-1/2,0,0,\lbrace q_{i}\rbrace}$ is also an eigenstate. All other remaining states can be expressed in terms of eigenstates of the Hamiltonian obtained from diagonalizing $H_{2}(\sigma,\nu,\alpha,J,M_{J})$ in Eq.~(\ref{app:eq:H2}). With this, we have calculated the partition function $Z=Z(\sigma,\nu,\alpha,K)$, which allows us to find the thermal expectation value of the electron spin according to Eq~(\ref{eq:avSz}). In order to confirm that there is still a spin flip in the presence of the flip-flop terms, we calculated this thermal expectation for up to $K=60$ nuclear spins. In Fig. \ref{app:fig:Sz_0_K}, we show the electron spin at zero temperature. 
\begin{figure}
\centering
\includegraphics[width=0.98\linewidth]{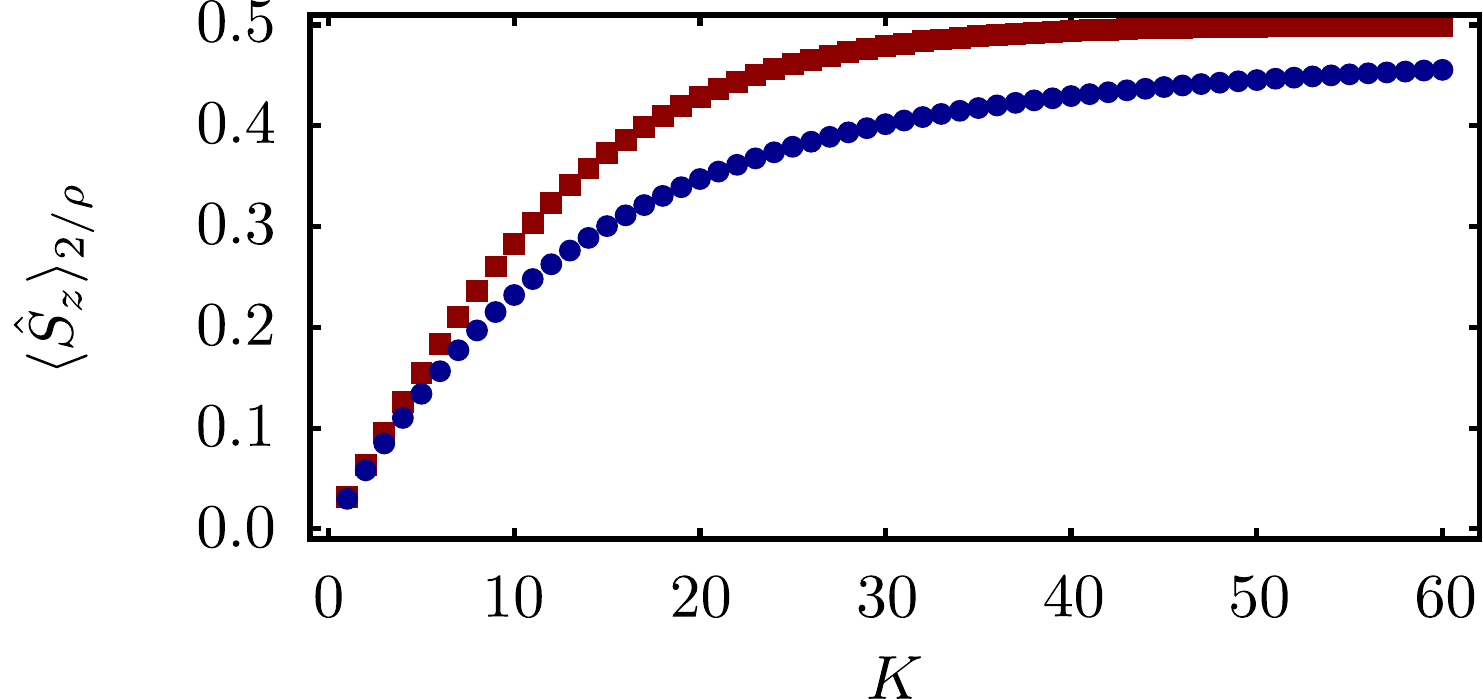}
\caption{(Color online)  Dependence of the thermal expectation value of the electron spin $\av{\oS_{z}}_{\tau}$ at $\tau=2/\rho$ on the number of nuclear spin with (blue circles) and without (red squares) the flip-flop terms. At this temperature the electron reaches its maximum. The remaining parameters are $\rho=0.35$, $\nu=2$, and $\sigma=K\rho\nu$.}
\label{app:fig:Sz_rho_K}
\end{figure}
Clearly, the flip-flop terms reduce the polarization of the electron spin for small system sizes. For larger $K$, the thermal expectation value $\av{\oS_{z}}_{0}$, however, tends to the same value as without the flip-flop terms. In the next section, we analytically show, that $\av{\oS_{z}}_{0}=-1/2$ is exactly reached for large system sizes. Moreover, we have analyzed the maximum value of the electron spin $\av{\oS_{z}}_{2/\rho}$, which is approximately found at $\tau=2\rho$, c.f. Fig. \ref{fig:pi_sz}. As is clear from Fig. \ref{app:fig:Sz_rho_K}, this quantity tends to $\av{\oS_{z}}_{2/\rho}=1/2$ for a large number of nuclear spins $K$. Note, that also for $\alpha=0$, this quantity is a function of $K$. Since the quantum fluctuations due to the flip-flop terms are already at zero temperature irrelevant for large system sizes, it is evident that they are also unimportant at higher temperatures.

\section{Ground state of the Hamiltonian}
\label{app:sec:ground_state}
In this section, we want to analytically find the ground state of the full Hamiltonian $\oH_{1}$. We choose $\sigma=K\rho\nu$, $\rho<1$ and $\rho\nu<1$, which are the necessary conditions for the spin flip. Moreover, we assume the Zeeman energy of a single nuclear spin to be $\nu>1$, which can be always reached by a suitable choice of the external magnetic field in an experiment. In order to find the ground state energy, we have first identified the minimum of the energies $E_{2,-}(\sigma,\nu,1,J,M_{J})$ for a fixed $J$ with respect to $M_{J}$. From this set of minima, we have then found the smallest energy with respect to $J$. Finally, we compare this minimum to the energies $E_{1}(\sigma,\nu,J,m_{S})$ resulting from the one-dimensional sub-spaces of the Hamiltonian. Following this scheme, we have found the ground state energy to be given by $E_{2,-}(\sigma,\nu,1,\frac{K}{2},\frac{K}{2})$ defined in Eq.~(\ref{app:eq:E2}). The corresponding ground state is given by
\begin{align}
&
\ket{E_{2,-}(K\rho\nu,\nu,\alpha,J,M_{J})}=
\nonumber\\
& \phantom{+\;\,}
\eta^{\frac{1}{2}}_{-}(K\rho\nu,\nu,\alpha,J,M_{J}) \ket{\frac{1}{2},J,M_{J}-1}
\nonumber\\
& +
\eta^{-\frac{1}{2}}_{-}(K\rho\nu,\nu,\alpha,J,M_{J}) \ket{-\frac{1}{2},J,M_{J}}
\,,
\end{align}
where
\begin{equation}
\eta^{-\frac{1}{2}}_{-}(K\rho\nu,\nu,1,\frac{K}{2},\frac{K}{2})
=
\frac{r-s}{t}\frac{1}{[(\frac{r-s}{t})^2+1]^{\frac{1}{2}}}
\end{equation}
and
\begin{equation}
\eta^{\frac{1}{2}}_{-}(K\rho\nu,\nu,1,\frac{K}{2},\frac{K}{2})
=
\frac{1}{[(\frac{r-s}{t})^2+1]^{\frac{1}{2}}}
\,,
\end{equation}
with $r=1-K-\nu+K\rho\nu$, $t=2K^{\frac{1}{2}}$ and, $s=(r^2+t^2)^{\frac{1}{2}}$. In the limit of large $K$, $\eta^{-1/2}_{-}(K\rho\nu,\nu,1,\frac{K}{2},\frac{K}{2})=1$ and $\eta^{1/2}_{-}(K\rho\nu,\nu,1,\frac{K}{2},\frac{K}{2})=0$ and, hence, also for the full Hamiltonian $\oH_{1}$ the ground state exhibits $\av{\oS_{z}}_{0}=-1/2$.

\end{document}